\documentclass[10pt]{article}
\usepackage{geometry}
\usepackage{xcolor}
\definecolor{graycolor}{gray}{0.9} 
\usepackage{microtype}
\usepackage{setspace} 
\usepackage[utf8]{inputenc}
\usepackage[english]{babel}
\usepackage{times}
\usepackage{array}
\usepackage{soul}
\usepackage{cite}
\usepackage[numbers,sort&compress]{natbib}
\usepackage{natbib}

\usepackage{titlesec} 
\titleformat {\section} [block] {\raggedright \fontsize{10}{10}\selectfont\bfseries} {\thesection. \space} {0pt} {}
\titlespacing {\section} {0pt} {12pt} {6pt}
\titleformat {\subsection} [block] {\raggedright \fontsize{10}{10}\selectfont\itshape} {\thesubsection .\space} {0pt} {}
\titlespacing {\subsection} {0pt} {12pt} {6pt}
\titleformat {\subsubsection} [block] {\raggedright \fontsize{10}{10}\selectfont} {\thesubsubsection .\space} {0pt} {}
\titlespacing {\subsubsection} {0pt} {12pt} {6pt}
\titleformat {\paragraph} [block] {\raggedright \fontsize{10}{10}\selectfont} {} {0pt} {}
\titlespacing {\paragraph} {0pt} {12pt} {6pt}

\usepackage{array} \newcommand{\PreserveBackslash}[1]{\let\temp=\\#1\let\\=\temp}
\newcolumntype{C}[1]{>{\PreserveBackslash\centering}m{#1}}
\newcolumntype{R}[1]{>{\PreserveBackslash\raggedleft}m{#1}}
\newcolumntype{L}[1]{>{\PreserveBackslash\raggedright}m{#1}}

\usepackage{lineno}
\usepackage{tabularx}
\usepackage{colortbl}
\usepackage{graphicx}
\usepackage{float}
\usepackage[export]{adjustbox}
\usepackage{caption}
\usepackage{fancyhdr} 
\usepackage{bm}

\usepackage{amsmath}
\usepackage{lastpage}
\usepackage{layout}
\usepackage{setspace} 
\usepackage{enumitem}
\usepackage{booktabs}
\usepackage{arydshln}
\usepackage{multirow}
\usepackage{color}
\usepackage{hyperref} 
\hypersetup{
	colorlinks=true,
	linkcolor=blue,
	filecolor=blue,
	urlcolor=black,
	citecolor=cyan,
}

\setcitestyle{open={[},close={]},citesep={,\!},numbers}

\fancypagestyle{firstpage}{
    \setlength{\headsep}{2.2cm}
    
    \setlength{\footskip}{1.5cm}
    \fancyhf{}
    \lhead{\begin{table}[H]
        \centering
        \begin{tabular}{L{2.5cm}C{10cm}C{3.1cm}R{2cm}}
            \includegraphics[scale=0.035]{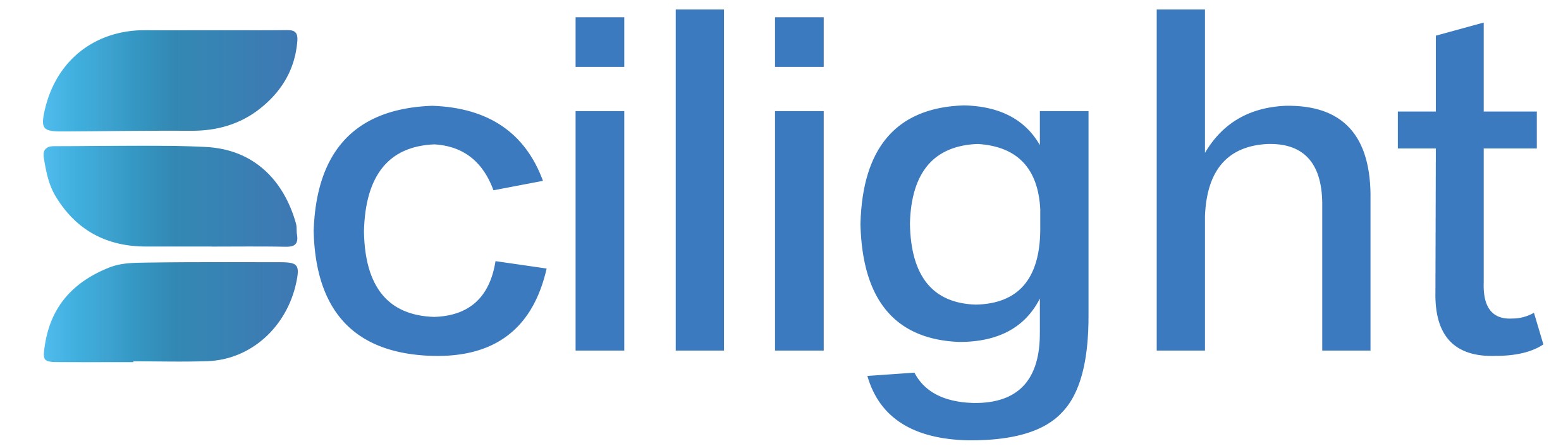} \vspace{-6pt}& \cellcolor{graycolor}\begin{tabular}[c]{@{}c@{}}\textit{International Journal of Gravitation and Theoretical Physics}\\ \href{https://www.sciltp.com/journals/ijgtp}{https://www.sciltp.com/journals/ijgtp}\end{tabular} & \includegraphics[width=1.15cm]{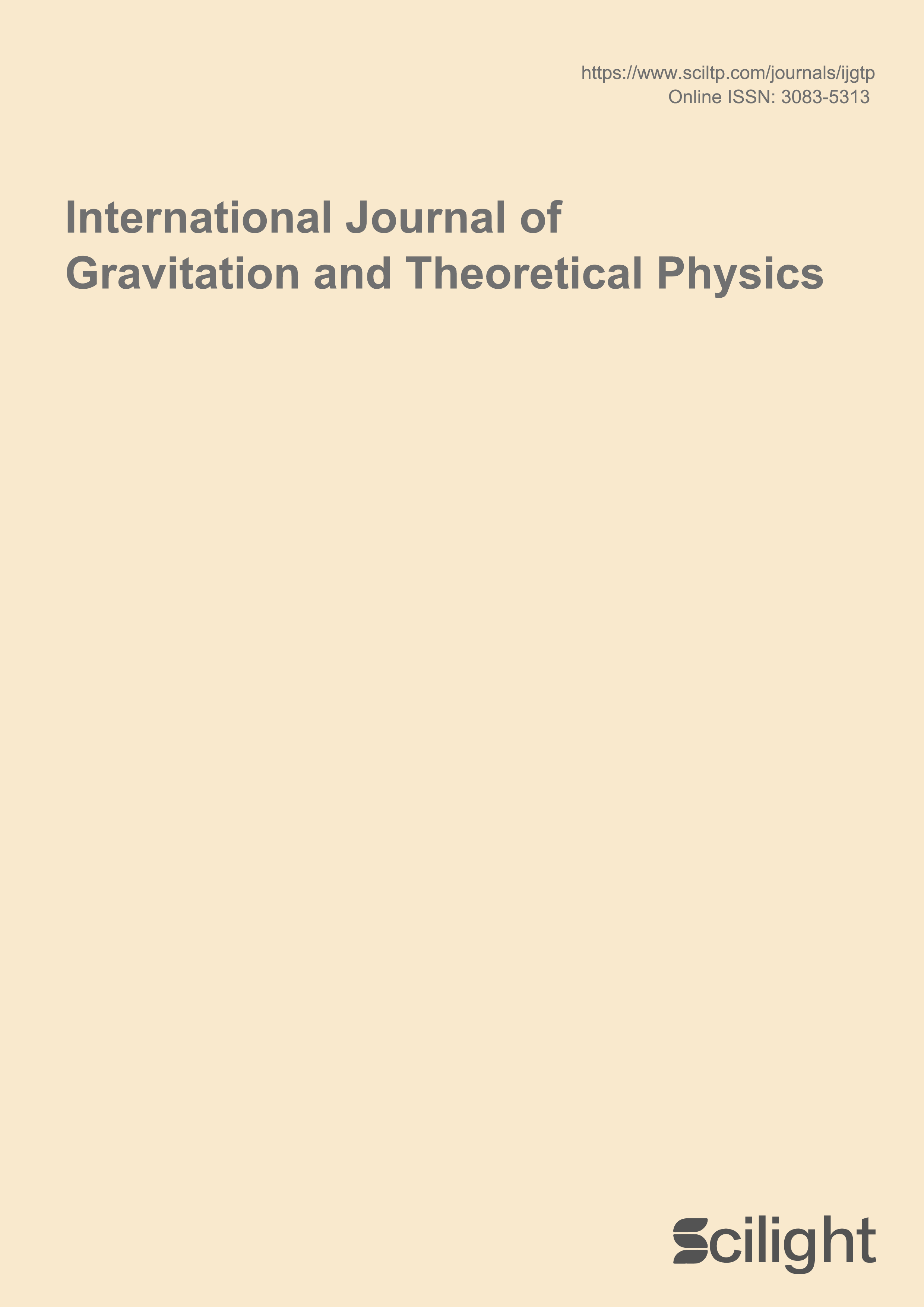} \vspace{-3pt}\\
        \end{tabular}
        \vspace{-22pt}
    \end{table}}
   
    \fancyfoot[C]{
        \vspace{-1.55cm}
        \begin{table}[H]
            \begin{minipage}[c]{0.15\columnwidth}
                \includegraphics[scale=0.5]{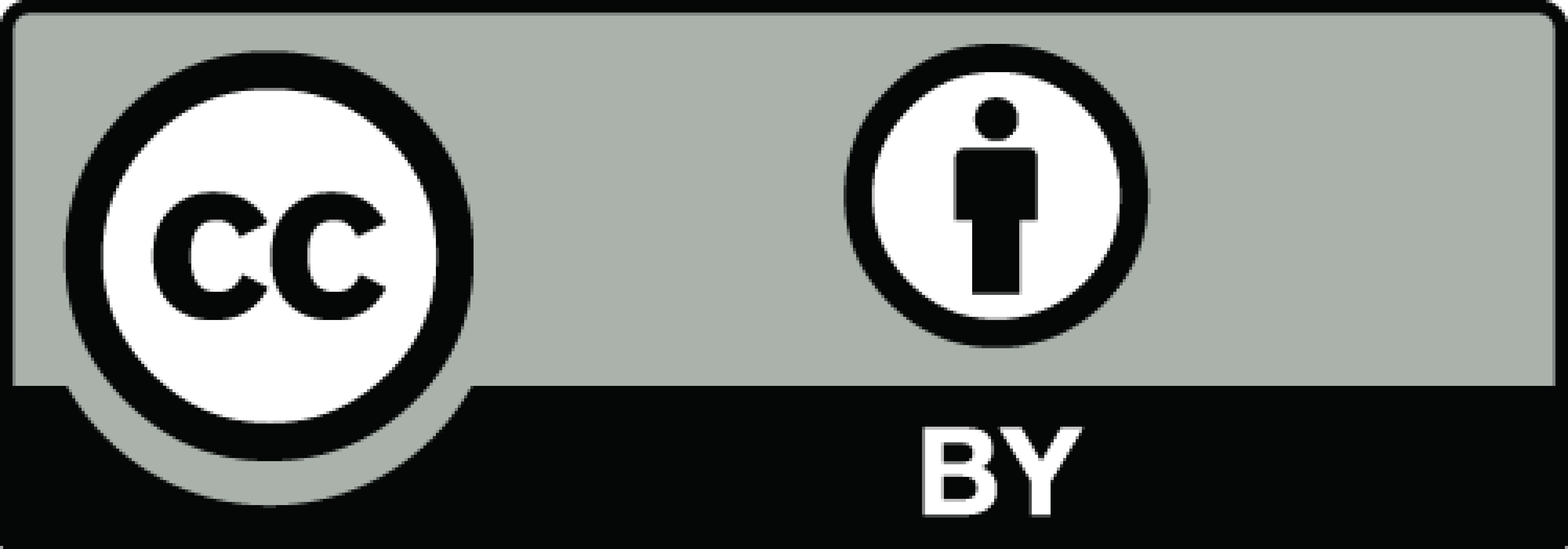} \vspace{1.1pt}
            \end{minipage}
            \hfill
            \begin{minipage}[c]{0.85\columnwidth}
                \scriptsize \textbf{Copyright:} © 2025 by the authors. This is an open access article under the terms and conditions of the Creative Commons Attribution (\mbox{CC BY}) license  (\href{https://creativecommons.org/licenses/by/4.0/}{https://creativecommons.org/licenses/by/4.0/}). \\ \textbf{Publisher’s Note:} Scilight stays neutral with regard to jurisdictional claims in published maps and institutional affiliations.
            \end{minipage}
    \end{table}}
    \vspace{-0.55cm}
}

\begin{document}
\newgeometry{left=2.5cm, right=2.5cm, top=1.8cm, bottom=4cm}
	\thispagestyle{firstpage}
	\nolinenumbers
	{\noindent \textit{{Review}}}
	\vspace{4pt} \\
	{\fontsize{18pt}{10pt}\textbf{Black Holes in Asymptotic Safety: A Review of Solutions and Phenomenology}}
	\vspace{16pt} \\
	{\large Andrea Spina \textsuperscript{1,2} }
	\vspace{6pt}
	 \begin{spacing}{0.9}
		{\noindent \small
			\textsuperscript{1}	\parbox[t]{0.98\linewidth}{INFN, Sezione di Catania, via Santa Sofia 64, 95123 Catania, Italy; Andrea.spina@phd.unict.it } \\ 
			\textsuperscript{2}	Department of Physics and Astronomy, Università di Catania, via Santa Sofia 64, 95123 Catania, Italy  			
\vspace{6pt}\\
		\footnotesize	\textbf{How To Cite}: Spina, A. Black Holes in Asymptotic Safety: A Review of Solutions and Phenomenology. \emph{International Journal of Gravitation and Theoretical Physics} \textbf{2025}, \emph{1}(1), 8. \href{https://doi.org/10.53941/ijgtp.2025.100008}{https://doi.org/10.53941/ijgtp.2025.100008}}\\
	\end{spacing}

\begin{table}[H]
\noindent\rule[0.15\baselineskip]{\textwidth}{0.5pt} 
\begin{tabular}{lp{12cm}}  
 \small 
  \begin{tabular}[t]{@{}l@{}} 
  \footnotesize  Received: 19 September 2025 \\
  \footnotesize  Revised: 30 September 2025 \\
   \footnotesize Accepted:  9 October 2025 \\
  \footnotesize  Published:  15 October 2025
  \end{tabular} &
  \textbf{Abstract:} Asymptotic Safety offers a conservative and predictive framework for quantum gravity, based on the existence of a renormalization group fixed point that ensures ultraviolet completeness without introducing new degrees of freedom. Black holes provide a natural arena in which to explore the implications of this scenario, as they probe the strongest gravitational fields and highlight the shortcomings of classical general relativity. In recent years, a variety of quantum-corrected black-hole solutions have been constructed within the Asymptotic Safety approach, either by renormalization-group improvement of classical metrics or through effective actions inspired by the flow of couplings. This review summarizes the current status of these developments. We discuss the structure and properties of the proposed solutions, their thermodynamics and evaporation, and their dynamical aspects such as quasinormal modes and shadows.\\
\\
  & 
\end{tabular}
\noindent\rule[0.15\baselineskip]{\textwidth}{0.5pt} 
\end{table}

	\section{Introduction}\label{Sec1}

Black holes provide a unique laboratory for testing gravity in its most extreme regime. Within classical general relativity, their geometry, stability, and thermodynamics are well understood, yet the persistence of singularities and the breakdown of predictability at small scales signal the necessity of a quantum theory of gravity. Among the promising approaches to quantum gravity, the Asymptotic Safety (AS) \cite{Reuter:2019byg,Reuter:2001ag} program offers a minimal and conservative framework in which ultraviolet completeness can be achieved without introducing new degrees of freedom beyond those of Einstein gravity. This is realized through the existence of a non-trivial renormalization group fixed point, first proposed by Weinberg \cite{Weinberg:1979}, which ensures the predictivity of the theory at \mbox{arbitrarily high energies.}

In recent years, a wide variety of black-hole solutions have been studied in the AS framework. These include quantum-corrected Schwarzschild and Kerr metrics \cite{Koch:2014cqa,Platania:2023srt,Saueressig:2015xua}, regularized spacetimes with modified near-horizon structure, and effective geometries derived from improved renormalization-group flows. Phenomenologically, such solutions have been analyzed in terms of their horizons, thermodynamic properties, stability, and dynamical features such as quasinormal modes, late-time tails, and Hawking evaporation. Moreover, the connection between AS-inspired black holes and astrophysical observables---from shadows and accretion physics to gravitational-wave signals---has attracted significant interest, especially in light of recent progress in black-hole imaging and gravitational-wave astronomy \cite{LIGOScientific:2016aoc,EventHorizonTelescope:2019dse}.

The purpose of this review is to provide a coherent survey of the current state of research on a subset of black holes in the Asymptotic Safety scenario that have been studied in connection with astrophysical features, summarizing the various solution-generating techniques, the main physical features of the geometries obtained, and the phenomenological implications that arise. Our discussion aims both to highlight the common structures that 

\restoregeometry

\noindent emerge across different approaches and to indicate open problems where further progress is needed.

The review is organized as follows. Section~\ref{Sec2} provides a brief overview of the Asymptotic Safety program, with emphasis on the renormalization group techniques relevant for black-hole physics and summarizes the various approaches that have been employed to construct quantum-corrected black-hole spacetimes, including RG-improved solutions and effective actions. Sections~\ref{Sec3}--\ref{Sec5} are devoted to the phenomenology of these geometries, covering  quasinormal modes (Section~\ref{Sec3}), basic thermodynamic properties (Section~\ref{Sec4}) and shadows cast by these black holes (Section~\ref{Sec5}).  Finally, Section~\ref{Sec6} contains our conclusions and outlines future directions \mbox{of research in this area.}

\section{Asymptotic Safety Framework}\label{Sec2}

The idea that gravity might be non-perturbatively renormalizable by approaching a non-Gaussian fixed point at high energies was originally proposed by Weinberg under the name of asymptotic safety \cite{Weinberg:1979}. This concept was made concrete in the functional renormalization group (FRG) framework by Reuter \cite{Reuter:1996cp}. In this approach, the central object is the effective average action $\Gamma_k$, a scale-dependent effective action constructed via a Wilsonian \cite{Wilson:1973jj} coarse-graining procedure. As first introduced  to gravity by Reuter \cite{Reuter:1996cp}, $\Gamma_k$ smoothly interpolates between the classical (bare) action $S$ in the ultraviolet ($k \to \infty$) and the full effective action $\Gamma$ in the infrared limit ($k\to 0$).
{The scale parameter $k$ acts as an infrared cutoff: fluctuations with momenta $p^2 > k^2$ have already been integrated out, while long-wavelength fluctuations with $p^2 < k^2$ are suppressed. In this sense, $\Gamma_k$ describes a renormalization-group trajectory in the space of effective actions, interpolating from microscopic to macroscopic physics and} its evolution with the scale $k$ allows a non-perturbative investigation of the theory space. {The physical identification of $k$ in curved spacetimes is not unique, but there are some general criteria which need to be respected, as will be discussed below.} The most common approach is to explore the evolution of $\Gamma_k$ with an exact functional renormalization group equation, which reads \cite{Wetterich:1992yh,Reuter:1996cp,Wetterich:1989xg}:
\begin{equation}
   k \partial_k\Gamma_k=\frac{1}{2}Tr\left(\frac{k \partial_{k}R_k}{\Gamma_{k}^{(2)}+R_k}\right),
\end{equation}
where $\Gamma_{k}^{(2)}$denotes the second functional derivative with respect to the fields, and $R_k$ is the IR regulator function. This equation is exact and non-perturbative, and its solutions define a trajectory in the space of action functionals. An alternative approach employs, the proper time (PT) version of the flow equation \cite{deAlwis:2017ysy,Bonanno:2004sy}, although it does not formally qualify as an exact flow equation for the effective average action, the proper time approach has been reconsidered in recent work \cite{Bonanno:2019ukb,Bonanno:2025tfj} as potentially capturing key aspects of Wilsonian coarse-graining.

A theory is said to be asymptotically safe if this trajectory approaches a non-Gaussian fixed point (NGFP) as $k\to\infty$ where all couplings remain finite and the theory is predictive. This framework has been applied to a variety of physical systems, including cosmology \cite{Zholdasbek:2024pxi,Bonanno:2024xne,Platania:2020lqb,Bonanno:2017kta,Bonanno:2016dyv} and black hole physics \cite{Koch:2014cqa,Platania:2023srt,Bonanno:2000ep,Held:2019xde,Platania:2019kyx,Bonanno:2024wvb,Bonanno:2023rzk,Bonanno:2025dry}. In the following, we focus on the application of the asymptotic safety scenario to the study of black hole solutions.

\subsection{Static Black Holes from RG Improvement}\label{Sec2.1}

The first black hole solution within the Asymptotic Safety framework was developed (by Bonanno and Reuter in Ref. \cite{Bonanno:2000ep}) using a procedure known as RG improvement. This method consists of substituting the classical couplings in the equations of motion or in the solutions with their scale-dependent counterparts obtained from the renormalization group flow. As outlined in the review \cite{Platania:2023srt}, the standard approach begins with a truncation of the effective average action—typically starting with the simplest case, the Einstein–Hilbert truncation—in which the couplings (such as Newton’s constant and the cosmological constant) are promoted to running quantities.
\begin{equation}
    \Gamma_k=\frac{1}{16{\pi}G(k)}\int{d^{4}x\sqrt{-g}(R-2\Lambda(k))},
\end{equation}

From this ansatz, the scale dependence of the couplings is obtained by solving the corresponding beta functions. For instance, the beta function for the dimensionless Newton's constant $g_k=k^2 G_k $ takes the form:
\begin{equation} \label{beta}
    k \frac{\partial g_{{k}}}{\partial k} = \beta_g \equiv \left(2 + \eta_N\right) g_k,
\end{equation}   
{where $\eta_N$ denotes the anomalous dimension, defined as:
$$\eta_N(g_k)=\frac{ B_1 g_k}{1- B_2 g_k},$$ 
with $B_1$ and $B_2$ that are some constants determined by the truncation \cite{Bonanno:2000ep}}. By integrating the flow equation for the beta function, typically derived using the background gauge field methods, one obtains an explicit expression for the running Newton’s constant \cite{Bonanno:2000ep,Platania:2023srt,Koch:2014cqa}
\begin{equation} \label{running G}
    G(k)=\frac{G_0}{1+{\omega}G_{0}k^2},
\end{equation}
where $G_0$ is the low energy Newton constant and $\omega$ is a constant which depends on the value of the fixed point \cite{Bonanno:2000ep}.

{We now apply the RG improvement procedure to classical black-hole solutions.} In this review we considered only spherically symmetric spacetime of the form 
\begin{equation} \label{line element}
  ds^2=-f(r)dT^2+\frac{1}{f(r)}dr^2+{r^2}d\Omega^2.  
\end{equation}
with $f(r)=1-\frac{2G(r)M}{r}$, where we have to substitute the classical gravitational constant with his running counterpart, which contains the deviations from general relativity.

{A key point concerns the identification of the physical scales of the system with the cutoff scale $k$.  This identification allows $k$ to be understood as a physical IR cutoff associated with a specific length scale of the system.} There are some general criteria that any physically meaningful identification of the cutoff scale $k(r)$ is expected to fulfill \cite{Koch:2014cqa}. First, the choice of $k(r)$ should be coordinate-independent, meaning that it must be constructed from diffeomorphism-invariant quantities, such as proper distances or curvature scalars, so as to preserve the geometric nature of the theory. Lastly, the functional form of $k(r)$ should be compatible with the symmetries of the underlying spacetime geometry. {Different choices of cutoff identification therefore give rise to} different improved solutions.

\subsubsection{Bonanno-Reuter {Metric}}\label{Sec2.1.1}
{The first solution we consider is also the first historically developed, namely the Bonanno–Reuter model \cite{Bonanno:2000ep}}, further explored in \cite{PhysRevD.73.083005}. In analogy with QED theory, where the Wilsonian infrared cutoff scale \(k\) is typically identified with the inverse of the radial coordinate, $(k \sim 1/r)$, Bonanno and Reuter proposed in \cite{Bonanno:2000ep,Bonanno:1998ye} a scale-setting procedure based on proper distances. In their approach, the RG scale $k(r)$ is linked to the proper radial distance $d(r)$ from the center $r = 0$ to a generic point at radius $r$ along a radial geodesic, such that
\begin{equation}
    k(r) \sim \frac{\xi}{d(r)},
\end{equation}
where $\xi$ is a numerical constant that sets the scale of quantum gravity effects. $d(r)$, in particular, {is obtained by interpolating the behavior between the} small and large radii of the radial proper distance, and reads:
\begin{equation}
    d(r)=\sqrt{\frac{2r^3}{2r+9G_0 M}}.
\end{equation}

This function for small radii goes as $\sim r^{3/2}$ while for large distances reduces to the classical $\sim r$. {Using this relation between $k$ and the position, one can construct the RG-improved metric function $f(r)$}:
\begin{equation} \label{B-R}
    f(r)=1-\frac{2MG_{0}r^2}{r^3+\tilde{\omega}{G_0}(r+\frac{9}{2}{G_0}M)}.
\end{equation}
where $\tilde{\omega}=\omega \xi^2$. Considering the correction to the Newtonian potential obtained from this framework \cite{Bonanno:2000ep}, one can fix the value of $\tilde{\omega}=118/15\pi$ to be in agreement with perturbative quantization of Einstein gravity \cite{Donoghue:1993eb,Hamber:1995cq}.

The metric function given in Equation \eqref{B-R} asymptotically approaches the Schwarzschild solution as $ r \to \infty$ (as do all the other solutions discussed in this review), while in the limit $ r \to 0 $ it exhibits a regular de Sitter core, thus avoiding the curvature singularity \cite{Bonanno:2000ep,Platania:2023srt}. 

As in many other scenarios (including all the models considered in this review), the horizon structure depends on the value of a free parameter. In particular, one typically finds three possible configurations: two horizons, a degenerate (extremal) horizon, or no horizon at all, depending on whether the parameter lies above, at, or below a critical threshold. In the Bonanno-Reuter (B-R) solution, the only free parameter is the black hole mass. Below a certain critical mass, the geometry becomes horizonless. {In the following, adopting Planck units ($G_0 = c = \hbar = 1$), so that lengths and masses are measured in multiples of the Planck length 
$$  \ell_{ Pl} = \sqrt{\frac{G_0\hbar}{c^3}} \,,$$
and the Planck mass 
   $$M_{Pl} = \sqrt{\frac{\hbar c}{G_0}} \,,$$}
 the critical value is approximately \cite{Bonanno:2000ep,Konoplya:2022hll}:
$$ M_{\text{crit}} \simeq 3.503. $$

\subsubsection{Hayward {Metric}}\label{Sec2.1.2}
{As discussed earlier,} the functional form of $k(r)$ is not unique and establishes different metric solutions. A reasonable choice could come from curvature scalars, particularly, in Ref. \cite{Held:2019xde}, they have connected $k$ with the Kretschmann scalar, which for classical case is $K=48 G_0^2 M^2 /r^6$ and they choose
\begin{equation}
    k^2=\alpha K^{1/2}.
\end{equation}
with the appropriate choice of $\alpha$ and {measuring the mass in units of length,} one can rewrite the metric  as:
\begin{equation}
    f(r)=1-\frac{2r^2/M^2}{r^3/M^3 + \gamma}
\end{equation}
where {$\gamma$ encodes both the non-universal value of the fixed point and the parameter $\alpha$, thereby providing} the quantum correction to the metric.

It is straightforward to see that this is the form of the notorious Hayward metric \cite{Hayward:2005gi}, which was obtained from simpler argumentation and not correlated with the AS or FRG and describes a quite simple regular black hole.

As the previous case, we recover the Schwarzschild solution asymptotically, {while varying the parameter determines the horizon structure}. Specifically, the critical value is 
$$\gamma_{crit}\simeq \frac{32}{27}$$
{below which the solution describes a black hole with two horizons}.

\subsubsection{Dymnikova {Metric}}\label{Sec2.1.3}

The scale identification $k(r)$ which links the deviation from classical spacetime behavior to the running of the gravitational couplings under the renormalization group flow, present some criticalities, When computing physical quantities such as the proper distance or the Kretschmann scalar, one typically relies on the classical background metric $\bar{g}_{\mu\nu} = g^{(0)}_{\mu\nu}$. However, this classical description breaks down at very short distances ($r \ll \ell_{\text{Pl}}$), precisely where quantum gravitational corrections are expected to dominate. Instead, by evaluating these invariants on the improved geometry $g^{(1)}_{\mu\nu}$—which incorporates running couplings via the renormalization group (RG)—one obtains different results, reflecting quantum effects on spacetime structure.

Moreover, in gravity, the identification of the cutoff scale $k(x)$ is not independent of the coupling being improved. Since $k(x)$ is defined using geometric quantities, which themselves depend on the metric, and hence on $G_0$, a nontrivial feedback occurs. This implies that the improvement procedure influences the spacetime in a way that affects the very definition of $k(x)$, leading to potential backreaction effects.

To address this interdependence, one can implement the improvement recursively \cite{Platania:2019kyx}: define a sequence of metrics $g^{(n)}_{\mu\nu}$ where, at each iteration, the running coupling $G_n(r)$ is determined through a cutoff $k_n(x)$ that depends on the previous geometry, i.e., $k_n[g^{(n-1)}_{\mu\nu}]$. Through this iterative construction, a self-consistent quantum-corrected metric $g^\infty_{\mu\nu}$ may be obtained, accounting for the mutual dependence between the geometry and the RG flow.

We begin by considering the classical Schwarzschild solution, described by the standard spherically symmetric line element \eqref{line element}. This geometry satisfies the vacuum Einstein field equations,
\begin{equation}
    R_{\mu\nu} - \frac{1}{2} R\, g_{\mu\nu} = 0,
\end{equation}
and is characterized by the lapse function
\begin{equation}
    f^{(0)}(r) = 1 - \frac{2 m G_0}{r}, \label{eq:Schwarzschild_f}
\end{equation}
where $G_0$ denotes the classical (constant) Newton coupling. To incorporate quantum gravity effects via the renormalization group (RG), we now perturb the system by promoting the Newton constant to a scale-dependent quantity, replacing $G_0$ with its running counterpart $G(k(r))$ given by Equation \eqref{running G}. The resulting quantum-corrected geometry can be interpreted as an exact solution of the Einstein field equations sourced by an effective energy-momentum tensor. This tensor takes the form
\begin{equation}
    T^{\text{eff}}_{\mu\nu} = (\rho + p)\left( \ell_\mu n_\nu + \ell_\nu n_\mu \right) + p\, g_{\mu\nu},
\end{equation}
where $\ell^\mu$ and $n^\mu$ are null vectors. The quantities $\rho$ and $p$ represent, respectively, an effective energy density and pressure, and arise from the radial variation of the Newton coupling $G(r)$ (see Ref. \cite{Platania:2019kyx} for details).

This energy-momentum tensor can be interpreted as the manifestation of quantum gravitational vacuum polarization. In this picture, $\rho$ corresponds to an effective self-energy contribution due to quantum gravitational effects. A small variation of $G(r)$ induces a modification in the background geometry, which in turn leads to further corrections to $G(r)$ through backreaction. Consequently, the energy density, $\rho \propto \partial_r \log G(r)$, can be regarded as a proxy for the strength of quantum effects within the black hole interior. This relation may be exploited in iterative schemes aimed at constructing a self-consistent cutoff identification $k^{(n+1)}(r)$ for $n > 1$.

The iterative procedure begins with the classical Schwarzschild geometry as the zeroth-order approximation. The first step corresponds to the replacement of the constant Newton coupling $G_0$ with its scale-dependent counterpart \eqref{running G}, as described earlier. 

Subsequent iterations, for $n > 1$, are defined by the recursive update
\begin{equation}
    G^{(n+1)}(r) = \frac{G_0}{1 + g^{-1} G_0\, k^{2}_{(n+1)}(r)},
\end{equation}
where the cutoff function $k^{(n+1)}(r)$ is constructed as a functional of the energy density $\rho^{(n)}(r)$ determined from the variation of $G^{(n)}(r)$ in the previous step.
\begin{equation}
    k^{2}_{(n+1)}(r) \equiv \mathcal{K}[\rho^{(n)}(r)].
\end{equation} 

Provided a specific choice for the functional $\mathcal{K}$ and under the assumption that the iterative procedure converges, one can explore the existence of a fixed-point configuration by considering the asymptotic limit $n \to \infty$. In this regime, the gravitational coupling becomes fully scale-dependent, yielding $G(r) \equiv G^{(\infty)}(r)$.

To constrain the structure of $\mathcal{K}$, Ref.~\cite{Platania:2019kyx} examines how classical curvature invariants—such as the Ricci scalar and the Kretschmann scalar—are affected by the spatial variation of Newton’s constant. {Requiring regularity at the center implies a balance between classical tidal forces and their quantum corrections, which guides} the construction of a self-consistent relation between the cutoff scale and the effective energy density \mbox{encoded in the running of $G(r)$.}

In particular, in the quantum-corrected curvature invariants, there is the appearance of the effective scale $G_0 \rho$, which arises naturally as a consequence of the spatial variation of the gravitational coupling. This quantity can be interpreted as a dynamical infrared cutoff, regulating the otherwise divergent classical invariants in the deep \mbox{interior region.}

Accordingly, a physically motivated identification for the RG scale is
\begin{equation}
k^2 \equiv \mathcal{K}[\rho] = \xi\, G_0 \rho,
\end{equation}
where $\xi$ is a positive, dimensionless parameter. This relation provides a concrete prescription to close the iterative improvement scheme by expressing the cutoff scale as a function of the effective energy density. Substituting the constrain, we obtain the metric function as \cite{Platania:2019kyx}:
\begin{equation}
    f(r)=1-\frac{2M}{r}\left(1-e^{-\frac{r^3}{2l^2 M}} \right).
\end{equation} 

{Here $l$ is defined as 
$$l=\sqrt{\frac{3\xi}{8\pi g_*}}\ell_{Pl},$$ 
where $\ell_{Pl}$ denotes the Planck length and $g_*$ is the non-trivial fixed-point value of the dimensionless Newton coupling}. This solution is a regular black hole of Dymnikova-type \cite{Dymnikova:1992ux}, with a de-Sitter core and which recover Schwarzschild, like the other solutions, for large radii. {Moreover, $l$ is a characteristic length scale of the order of the Planck length. Its critical value corresponds to the maximum scale for which an event horizon still exists; for larger $l$ the horizon disappears. Approximately, one finds} \cite{Konoplya:2023aph}:
\begin{equation}
    l_{\text{cr}} \approx 1.138\, M .
\end{equation}

\subsection{Dynamical Black Holes from Gravitational Collapse}\label{Sec2.2}

While the static RG-improved black hole solutions provide important insights into possible quantum-gravitational modifications of the spacetime geometry, they lack a dynamical process of black hole formation. To address this limitation, it is natural to consider more realistic scenarios involving gravitational collapse, where quantum effects are taken into account during the dynamical evolution of the matter content. This approach allows for a more physical modeling of black hole spacetimes, as it directly incorporates the quantum-gravity-improved dynamics leading to their formation.

In this context, one can regard these models as quantum-gravity-improved versions of the classical Oppenheimer-Snyder-Datt (OSD) \cite{Oppenheimer:1939ue,Datt:1938uwc} collapse scenario, where the RG flow is implemented dynamically. In recent years, interestingly, several works have developed such improved collapse models within the framework of asymptotic safety \cite{Bonanno:2023rzk,Harada:2025cwd,Bonanno:2025dry}. These studies aim to understand how the running of gravitational couplings can influence the causal structure and the formation of singularities, and whether quantum-gravity effects can lead to singularity resolution in a self-consistent way.

We focused on two correlated models that share the physical approach to obtain the new metric. Starting from the action of GR with matter \cite{Markov:1985py}:
\begin{equation}
\label{action}
S = \frac{1}{16 \pi G_N} \int d^4 x \sqrt{-g} \left[R + 2  \chi(\epsilon)  \mathcal{L}\right] .
\end{equation}
where, $\chi$ expresses a gravity–matter coupling function, which depends on the proper energy density $\epsilon$ of the matter fluid. The field equations {lead to a relation} between the matter sector and an effective running Newton constant (see \cite{Bonanno:2025dry,Harada:2025cwd,Bonanno:2023rzk} for details),
\begin{equation}
    8\pi G(\epsilon)=\frac{\partial (\chi \epsilon)}{\partial \epsilon}.
\end{equation}

Now, considering spherically homogeneous collapse, they assume a FLRW metric which is dependent on the scale factor $a(t)$
\begin{equation}
    ds^2 = -dt^2 + a^2(t) \left( \frac{dr^2}{1 - K r^2} + r^2 d\Omega^2 \right).
\end{equation}

In this framework, one can set $K = 0$ without loss of generality, as it can be shown that the final result obtained from the matching conditions does not depend on the specific value of $K$ \cite{Harada:2025cwd,Bonanno:2023rzk,Bonanno:2025dry}. Under this assumption, the modified field equations simplify to a Friedmann-like equation for the scale factor:
\begin{equation} \label{adot-pot}
    \dot{a}^2 = -V(a),
\end{equation}
where the effective potential $V(a)$ encodes the dynamics of the collapsing matter content. As discussed previously, this is related by the scale dependence of the Newton coupling, and the potential is defined as:
\begin{equation}\label{V(a)}
    V(a) = -\frac{a^2}{3} \int_0^{\epsilon(a)} G(s) \, ds,
\end{equation}
with $\epsilon(a)$ representing the energy density as a function of the scale factor. The presence of the running coupling $G(s)$ captures the leading quantum gravitational corrections to the classical collapse dynamics. {Once the functional form of $G(\epsilon)$ is specified, the full collapse dynamics can be derived.}  

As in OSD model it is supposed that this interior collapsing has to match a static and spherically symmetric spacetime outside which line element is:
\begin{equation}
  ds^2=-f(R)dT^2+\frac{1}{f(R)}dR^2+{R^2}d\Omega^2.  
\end{equation}

The matching condition, given by Israel \cite{Israel:1966rt}, are imposed to a boundary hypersurface at comoving radius $r=r_b$ and can be reduced to \cite{Malafarina:2022wmx}:
\begin{equation} \label{matching}
    1=f(R_b)+\dot{R_b}^2
\end{equation}
with $R_b(T(t))=r_ba(t)$. From Equation \eqref{matching} we can obtain the new solution of the black hole.

\subsubsection{Proper Time Approach}\label{Sec2.2.1}
To obtain the form of the running gravitational constant we will consider the asymptotic safety framework, hypothesizing that the collapse occurs within that scenario. Specifically, within the proper time approach described at the beginning of the section. They use a flow equation for the Wilsonian effective action $S_\Lambda$, regularized via the proper-time integral, given by the following expression \cite{Bonanno:2004sy,Bonanno:2025tfj}:
\begin{equation}
    \Lambda\partial_\Lambda S_\Lambda=\frac{1}{2}\int_0^\infty \frac{ds}{s} r_\Lambda(s)\mathrm{STr}\left( e^{-s \tilde{S}_\Lambda^{(2)}} \right).
\end{equation}

Here, $\Lambda$ is the momentum scale, $\tilde{S}_\Lambda^{(2)}$ is the hessian of the gauge fixed Wilsonian effective action and of the ghost action, and $r_\Lambda(s)$ contains the regulating function. The flow equation can be treated, even in the Einstein-Hilbert truncation, with different approaches. We can use different schemes of regularization, field parametrization (linear or exponential) and different choices of gauge-fixing. Considering these configurations, the beta {function for the running Newton constant can be expressed as}
\begin{equation} \label{beta2}
    \Lambda \partial_\Lambda g_\Lambda = \beta_g \equiv \left(2 + \eta_N\right) g_\Lambda,
\end{equation}
{with anomalous dimension} \cite{Bonanno:2025tfj,Bonanno:2025dry}
\begin{equation}
    \eta_N=-\frac{\Omega g_\Lambda}{1-\varepsilon \frac{\Omega}{2}g_\Lambda}
\end{equation}
{where $\Omega$ is a dimensionless constant encoding the dependence on} the gauge, parametrization, and the regulating parameter, while $\varepsilon$ express the {chosen} regularization scheme.

In Ref. \cite{Bonanno:2025dry}, remarkably, they found that the solutions are formally independent from $\Omega$, it changes only the non universal numerical value of the fixed point {$g_*$}. What distinguishes different solutions is the choice of the parameter $\varepsilon$. {In the following we will refer to two possible values: $\varepsilon=0$, corresponding to the ``C scheme'' which leads to the model of \cite{Bonanno:2023rzk}, and $\varepsilon=1$, corresponding to the ``B scheme'' underlying the model studied in \cite{Bonanno:2025dry}.}

\subsubsection{{Scheme B ($\varepsilon=1$)}}\label{Sec2.2.2}

{We first analyze the case corresponding to scheme B ($\varepsilon=1$), considered in \cite{Bonanno:2025dry}. Integrating the beta function \eqref{beta2}
 yields the following running Newton constant}
\begin{equation} \label{Run_G}
    G(\Lambda)=\frac{2g_*}{\Lambda^2+\sqrt{4g_*^2+\Lambda^4}}, 
\end{equation}
{where $g_*$ is the fixed point value}. To establish a connection between the RG scale {$\Lambda$} and the proper energy density, interpreted, as previously, as  a dynamical infrared cutoff \cite{Platania:2019kyx}, thereby anchoring the running couplings to the physical characteristics of the system, {one sets $\Lambda^2=q\epsilon$, with $q$ a free parameter controlling the strength of quantum corrections.}

{In this model the scale factor never vanishes, ensuring} geodesic completeness and avoiding the singularity of the classical OSD collapse. Applying the junction condition with Equation \eqref{matching}, the new metric takes the form:

\begin{equation}
    f(R)=\frac{3M^2+qR^4-M\sqrt{9M+q^2 R^6}}{qR^4}+\frac{2}{3}qR^2\text{Arctanh}\left(\frac{\left(q-\sqrt{q^2+\frac{9M^2}{R^6}}\right)R^3}{3M}\right)
\end{equation}

{As in previous cases, the Schwarzschild limit is recovered at large radii, and varying the parameter $q$ (which, with a redefinition, also absorbs the fixed point value), the usual horizon structure emerges}. In particular, the critical value, in unit of mass, is $q\simeq 1.37$, below this {threshold the spacetime} has no horizons. 

\subsubsection{{Scheme C ($\varepsilon=0$)}}\label{Sec2.2.3}

Taking in consideration the regularization scheme {C ($\varepsilon=0$)}, as discussed in \cite{Bonanno:2023rzk,Bonanno:2021squ,Bonanno:2004sy}, {the running constant takes the following form}
\begin{equation}
G(k) = \frac{G_0}{1+G_0 \Lambda^2/g_*},
\end{equation}
{where $g_*$ is the non-trivial fixed point value.} With similar choice of the cutoff, connecting { $\Lambda$ and the proper energy density with a parameter $\xi$, using Equation \eqref{matching}, the solution becomes}:
\begin{equation}
     f(r)=1-\frac{r^2}{3\xi}\log{(1+\frac{6M\xi}{r^3})}.
\end{equation}

Also in this model the scale factor never goes to zero, {Schwarzschild is recovered at large radii, and a critical parameter value determines the horizon structure. Specifically, in units of mass ($G_0=1$), the critical value is $\xi\simeq0.46$, which sets the maximum parameter value for which horizons still exist}.

\subsubsection*{Metric Solutions}

\begin{figure}[H]
    \centering
    \includegraphics[width=0.8\columnwidth]{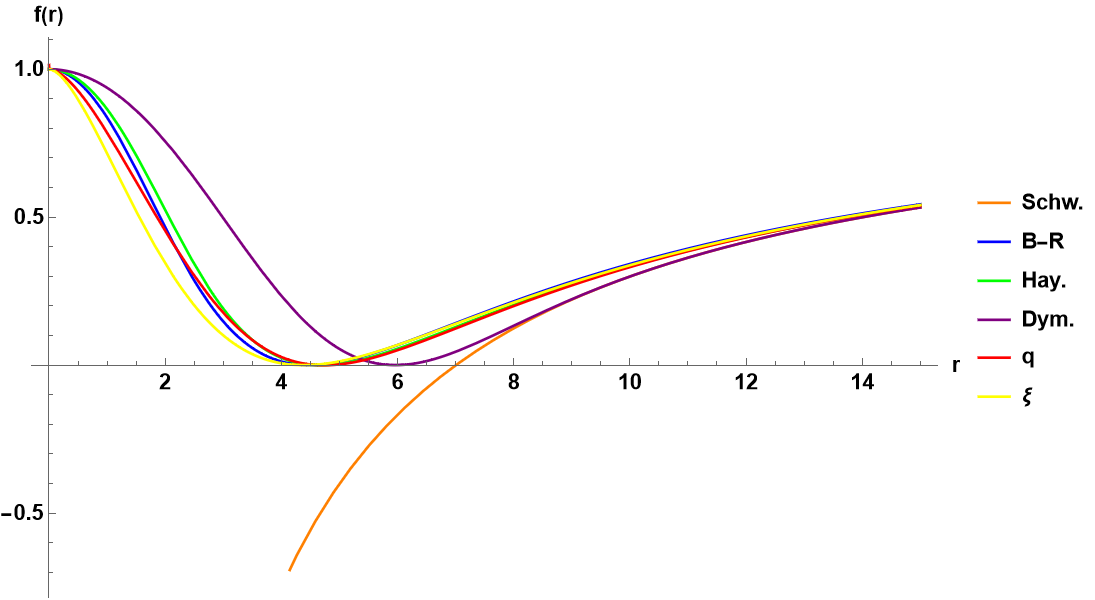}
    \caption{Metric functions $f(R)$ for all the regular black hole solutions at their critical values with $M=3.503$ (critical mass for B-R). }
\label{Metrics}
\end{figure}

After introducing all the solutions of interest, we proceeded to plot the corresponding metric functions $f(r)$ in order to analyze their behavior and perform a direct comparison, as shown in Figure~\ref{Metrics}. In particular, we focused on the cases close to the critical values for each model since it is the situation that differs most from the classical case, fixing the mass to $3.503$, which corresponds to the critical value for the Bonanno-Reuter (B-R) solution.

\vspace{-18pt}

As previously mentioned, we can observe that all the metric profiles converge to the Schwarzschild solution at large distances. However, a distinctive feature of these models inspired by Asymptotic Safety is that this convergence occurs already within a few horizon radii ($2–3r_H$).  {Thus, quantum corrections remain negligible at large scales} and become significant only in the near-horizon region, where high-energy effects and strong gravitational fields are expected to play a relevant role, precisely as one would anticipate in a quantum gravity framework.

{\section{Quasinormal Modes}\label{Sec3}}

{Having introduced the main classes of solutions, we now turn to their phenomenological properties, starting with quasinormal modes (QNM)}. Quasinormal modes are the characteristic oscillations of black holes and other compact objects, arising as solutions of the linearized perturbation equations subject to physically motivated boundary conditions \cite{Konoplya:2011qq,Berti:2009kk,Kokkotas:1999bd,Berti:2025hly}. For a stationary and asymptotically flat black hole, the perturbations can often be reduced to a master wave-like equation of the form
\begin{equation}
\frac{d^2 \Psi}{dr_*^2} + \left[\omega^2 - V(r)\right] \Psi = 0 ,
\end{equation}
where $r_*$ is the tortoise coordinate defined via $dr_* / dr = f(r)^{-1}$, $f(r)$ being the metric function, $V(r)$ is an effective potential depending on the background geometry and the spin of the perturbation which encodes the theory from which it comes and the nature of the perturbation, and $\omega$ is, in general, a complex frequency.

QNMs are defined by imposing purely ingoing waves at the event horizon and purely outgoing waves at spatial infinity:
\begin{equation}
\Psi(r_*) \sim e^{-i \omega r_*}, \quad r_* \to -\infty ,
\end{equation}
\begin{equation}
\Psi(r_*) \sim e^{+i \omega r_*}, \quad r_* \to +\infty .
\end{equation}

These boundary conditions reflect the absence of incoming radiation from either the horizon or infinity, corresponding to the natural ``ringdown'' response of the spacetime to an initial perturbation.

The complex frequency $\omega = \omega_R - i \omega_I$ encodes both the oscillation frequency ($\omega_R$) and the damping rate ($\omega_I$) of the mode. The QNM spectrum depends solely on the parameters of the black hole (mass, charge, spin, and possibly other model-specific quantities), making them an important tool for testing gravitational theories and the nature of compact objects, as widely used \cite{Daghigh:2020fmw,Bolokhov:2023ozp,Konoplya:2022hll,Dubinsky:2024gwo,Taylor:2024duw,Berti:2003jh}.

One crucial mathematical property is that QNM do not form a complete set of functions, unlike normal modes in closed systems. This incompleteness implies that the late-time evolution of a perturbation is not fully captured by a sum over QNMs. In the time domain, the ringdown signal is followed by a \textit{power-law tail} at asymptotically late times \cite{Price:1971fb,Price:1972pw,Konoplya:2025afm,Rosato:2025rtr,Konoplya:2023fmh,Koyama:2000hj}. The interplay between the QNM dominated phase and the late-time tail is essential for a complete description of black hole perturbations.

There are several semi-analytical and pure numerical methods to compute QNM frequencies \cite{Konoplya:2011qq,Leaver:1985ax,Nollert:1993zz,Konoplya:2019hlu,Jansen:2017oag}, {and results are available for most classical and modified-gravity solutions}. For the solutions and models we presented in the previous sections, QNM were calculated in several papers. {For the Bonanno–Reuter solution, test-field perturbations were first studied (with limited accuracy) in \cite{Rincon:2020iwy} and later refined in \cite{Konoplya:2022hll}, which also analyzed Hayward-type solutions. The QNM of the Dymnikova solution were computed in \cite{Konoplya:2023aph}, while for the collapse-based solutions the spectra were obtained in \cite{Stashko:2024wuq,Spina:2024npx,Bonanno:2025dry} for test fields. Gravitational perturbations were investigated} in \cite{Bolokhov:2025lnt} for B-R, in \cite{Bolokhov:2025egl,Malik:2025dxn} for Hayward, and in \cite{Bonanno:2025dry,Lutfuoglu:2025ohb} for the two proper-time solutions. For Dymnikova, only greybody factors were computed for gravitational perturbations \cite{Dubinsky:2025nxv}.

{Table \ref{QNM} summarizes the frequencies of the solutions close to their critical parameter values and compares them with the Schwarzschild case.} All these values were calculated in units of mass so {it cannot include the Bonanno-Reuter model.}. We report both the fundamental mode and the third overtone  {in order to highlight the peculiar behavior of the QNM spectrum.}
We can observe that, already for the fundamental mode, the imaginary part departs more significantly from the classical Schwarzschild case. In fact, all the models considered display both a larger imaginary part and a larger real part (with the exception of the Dymnikova solution, which instead shows the opposite behavior and, as can already be anticipated from the form of its metric, exhibits smaller deviations from Schwarzschild). As is well known, the fundamental mode is mostly determined by the shape of the effective potential, which in these modified metrics differs only slightly from the classical one, resulting in correspondingly small percent-level deviations.  

The situation changes for the overtones. It has been shown \cite{Konoplya:2022hll,Konoplya:2022pbc} that these modes are closely related to the structure of the event horizon, which, as illustrated in Figure \ref{Metrics}, departs significantly from the classical case. For this reason, the overtone spectrum is sometimes referred to as the distinctive ``sound of the event horizon'' \cite{Konoplya:2023hqb}. The phenomenon of increasing sensitivity of the higher overtones, leading to larger deviations from Schwarzschild, is often described as the ``outburst of overtones'' \cite{Konoplya:2022pbc}. Indeed, as shown in Table \ref{QNM}, already in the scalar case the third overtone displays a much more pronounced deviation in its real part.

\vspace{-6pt}

\begin{table}[H]
    \caption{Complex frequencies in unit of mass of scalar QNMs ($l=1$) for different regular black hole models, with parameters close to their critical values (the specific parameter choices are indicated in the table entries). Results are compared to the Schwarzschild case: $\omega_{\text{Schw}} = 0.29293 - 0.09766i$ for $n=0$ and $\omega_{\text{Schw}} = 0.203259 - 0.788298i$ for $n=3$. The table reports real and imaginary parts along with the corresponding relative percentage deviations.}
    \centering
    \begin{tabular*}{\textwidth}{@{\extracolsep{\fill}} c c c c c} 
        \toprule
        \textbf{$\bm{n=0}$} & \textbf{$\bm{\text{Re}(\omega)}$} & \textbf{Deviation from Schw. [\%]} & \textbf{$\bm{\text{Im}(\omega})$} & \textbf{Deviation from Schw. [\%]} \\
        \midrule
        Hayward {($\gamma=1$)}  & 0.30562 & $  4.32\%$ & 0.08559 & $ 12.37\%$ \\
        Dymnikova {($l=1.1$)} & 0.28874 & $ 1.43\%$ & 0.09468 & $ 3.05\%$ \\
        {Scheme B ($q=1.38$) }    & 0.30598 & $ 4.46\%$ & 0.08300 & $ 15.06\%$ \\
        {Scheme C ($\xi=0.455$)}    & 0.31052 & $ 6.01\%$ & 0.07986 & $ 18.22\%$ \\
        \midrule
        \textbf{$\bm{n=3}$} & \textbf{$\bm{\text{Re}(\omega)}$} & \textbf{Deviation from Schw. [\%]} & \textbf{$\bm{\text{Im}(\omega})$} & \textbf{Deviation from Schw. [\%]} \\
        \midrule
        Hayward   & 0.158106   & $ 22.21\%$ & 0.700610 & $ 11.12\%$ \\
        Dymnikova & 0.159      & $ 21.75\%$ & 0.964    & $ 22.33\%$ \\
        {Scheme B}  & 0.13019    & $ 35.93\%$ & 0.70745  & $ 10.24\%$  \\
        {Scheme C}  & 0.14249    & $ 29.91\%$ & 0.67509  & $ 14.34\%$ \\
        \bottomrule
    \end{tabular*}
    \label{QNM}
\end{table}

\section{Hawking Temperature}\label{Sec4}

The discovery of Hawking radiation in 1974 \cite{Hawking:1974rv,Hawking:1975vcx} established that black holes are not completely black, but instead emit a thermal spectrum of particles due to quantum field effects in curved spacetime. The original derivation was semiclassical in nature: gravity was treated as a fixed classical background, while quantum fields were quantized on top of it. Within this approximation, the radiation is almost perfectly thermal. When quantum gravitational corrections are included, however, deviations from the classical picture are expected, potentially altering the emission spectrum and the evaporation process, particularly in the extreme regime close to the endpoint of evaporation. These questions remain open and are an active line of research in quantum gravity.  

From a practical perspective, it is often convenient to retain a semiclassical treatment and study the evaporation of black holes whose geometry already incorporates quantum corrections. Furthermore, in the Schwarzschild case, it has been shown that the contribution of gravitons to the total massless particle emission is below the percent level, and becomes even less relevant when massive channels open at the final stages of evaporation \cite{Page:1976df,Page:1976ki}. This observation supports the use of the standard Hawking framework to estimate the flux of quantized test fields in a given corrected geometry, under the assumption that the graviton contribution is subdominant.  

In this approach, the black hole is considered to be in thermal equilibrium with its environment, so that its temperature remains constant between the emission of two successive particles. The system can therefore be described in the canonical ensemble, and the Hawking temperature takes the form \cite{Hawking:1975vcx}
\begin{equation}
    T_H = \frac{f'(r)}{4\pi}\Big|_{r=r_H},
\end{equation}
where $f(r)$ denotes the lapse function of the metric and $r_H$ is the event horizon radius. 

We now want to evaluate how the quantum corrections predicted within the Asymptotic Safety framework modify the Hawking temperature with respect to the classical Schwarzschild case, where in mass units the temperature reads $T_H^{\text{Schw}} = 1/(8\pi)$. The resulting values for the different regular black hole solutions considered here are summarized in Table \ref{HawkingT}, {with the near-critical values of the parameters used in the evaluation are explicitly specified in the table. These results} were obtained and compared in \cite{Konoplya:2023bpf} for Bonanno-Reuter ({see also \cite{Bonanno:2000ep,Platania:2023srt} for earlier analyses including the evaporating mass and the interpretation} of remnants), Hayward and Dymnikova spacetimes and in \cite{Bonanno:2025dry} for the non-singular collapse models.

\vspace{-6pt}

\begin{table}[H]
    \caption{Hawking temperature $T_{\mathrm{H}}$ for different regular black hole solutions evaluated near their critical values. Results are compared with the Schwarzschild case. As the critical configuration is approached, the solutions tend toward extremality and the temperature vanishes.}
    \label{HawkingT}
    \centering
    \newcolumntype{C}{>{\centering\arraybackslash}X}
    \begin{tabularx}{\textwidth}{C C}
        \toprule
        \textbf{Metric} & \textbf{$\bm{T_{\mathrm{H}}}$} \\
        \midrule
        Schwarzschild & $0.0398$ \\
        Bonanno-Reuter ($M = 3.505$) & $0.00071$ \\
        Hayward ($\gamma = 31/27$) & $0.00369$ \\
        Dymnikova ($l = 1.137$) & $0.00291$ \\
        Scheme B ($q = 1.38$) & $0.00472$ \\
        Scheme C ($\xi = 0.455$) & $0.00345$ \\
        \bottomrule
    \end{tabularx}
\end{table}

As can be seen {in Table \ref{HawkingT}}, all the regular black hole solutions display a Hawking temperature that is significantly suppressed with respect to the classical Schwarzschild case. In particular, {near the critical parameter values, where the inner Cauchy horizon approaches the event horizon, the surface gravity approaches zero and the solutions tend to extremal configurations, for which the Hawking temperature vanishes.}. This behavior implies that the corresponding evaporation process is drastically slowed down, leading to emission rates that are orders of magnitude smaller than in the classical scenario. From a phenomenological perspective, such a suppression may have important implications for the lifetime of primordial black holes and the possibility of long-lived remnants black holes. Moreover, despite quantitative differences among the models (which obviously depends on the exact value of the parameter considered close to the extremal configuration), the qualitative picture remains robust: quantum corrections systematically lower the Hawking temperature and drive the solutions toward cold, long-lived remnants.

\section{Shadows}\label{Sec5}

Motivated by the recent images obtained by the Event Horizon Telescope (EHT) \cite{EventHorizonTelescope:2019dse,EventHorizonTelescope:2022wkp} and the consequent rise in interest in the subject, another astrophysical feature we analyze for the regular black hole solutions within the Asymptotic Safety framework is their shadow. In order to determine the shadow of a solution in a spherically symmetric spacetime of the form \eqref{line element}, one first needs to compute the radius of the photon sphere. This corresponds to the existence of unstable circular null geodesics, which play a central role in defining the size and shape of the black hole shadow.

For null geodesics, the radial motion can be written in terms of an effective potential as \cite{alma9917223253502466,Chandrasekhar:1985kt}
\begin{equation}
\dot{r}^2 + V_{\text{eff}}(r) = E^2 , \qquad 
V_{\text{eff}} = f(r) \left(\frac{L^2}{r^2}\right),
\end{equation}
where $E$ and $L$ are the conserved energy and angular momentum of the photon. Circular null orbits are obtained by imposing $\dot r=0$ and $\ddot{r}=0$ and the extremality condition
\begin{equation}
\frac{d}{dr} V_{\text{eff}}(r) = 0 .
\end{equation}

This is equivalent to \cite{Cardoso:2008bp,Konoplya:2019sns}
\begin{equation}
\frac{d}{dr}\left(\frac{r^2}{f(r)}\right) = 0 ,
\end{equation}
which provides the photon sphere radius $r_p$.  

Finally, the shadow radius as seen by a distant observer is obtained from:
\begin{equation}
R_{\text{Shad}} = \frac{r_p}{\sqrt{f(r_p)}} .
\end{equation}

The dependence of the shadow size on the parameters of different regular black hole metrics has been investigated in various works: for the Bonanno--Reuter solution in \cite{Lambiase:2023hng}, for the {non-singular collapse model (scheme C)} in \cite{Stashko:2024wuq}, for Hayward-type black holes in \cite{Held:2019xde}, and for {the alternative dynamical model (scheme B)} in \cite{Bonanno:2025dry}. In Table \ref{Shadow} we report the values of the shadow radius computed close to the critical value of the respective parameters, in order to highlight the maximum deviation from the classical Schwarzschild case.

\vspace{-6pt}
\begin{table}[H]
    \caption{Shadow radius as seen from infinity for different regular black hole solutions evaluated close to their critical parameter values (explicitly indicated), compared with the classical Schwarzschild value. For reference, the Event Horizon Telescope (EHT) constrains the shadow size within the range $R_{\mathrm{Shad}} \sim 4.8M$--$5.2M$~\cite{Vagnozzi:2022moj,EventHorizonTelescope:2021dqv}.}
    \label{Shadow}
    \centering
    \newcolumntype{C}{>{\centering\arraybackslash}X}
    \begin{tabularx}{\textwidth}{C C}
        \toprule
        \textbf{Metric} & \textbf{$\bm{R_{\mathrm{Shad}}}$} \\
        \midrule
        Schwarzschild & $5.196$ \\
        Bonanno-Reuter ($M = 3.505$) & $4.83312$ \\
        Hayward ($\gamma = 31/27$) & $4.91693$ \\
        Dymnikova ($l = 1.137$) & $5.19599$ \\
        Scheme B ($q = 1.38$) & $4.94024$ \\
        Scheme C ($\xi = 0.455$) & $4.86582$ \\
        \bottomrule
    \end{tabularx}
\end{table}

Among the models considered, the Bonanno--Reuter solution exhibits the most pronounced reduction of the shadow size, while the Dymnikova metric remains nearly indistinguishable from Schwarzschild (as in the fundamental mode of QNM). The other models (Hayward, {non-singular collapse solutions}) display intermediate deviations, with shadow radii (for all the models considered) consistently lying within the range constrained by the EHT observations \cite{Vagnozzi:2022moj,EventHorizonTelescope:2021dqv}. We can also point out that in all the considered cases the quantum corrections tend to reduce the size of the shadow compared to the classical Schwarzschild solution. However, given the current precision of the EHT measurements, we cannot exclude any of these AS-inspired models nor place meaningful constraints on their parameters based on the shadow size alone.

\section{Conclusions}\label{Sec6}

Asymptotic Safety provides a compelling and conservative framework in which black-hole physics can be studied consistently at both classical and quantum levels. A large body of work has now established a variety of effective black-hole geometries obtained through renormalization-group improvements or more systematic derivations, and their properties have been investigated from multiple perspectives. These include horizon structure and singularity resolution, modifications to black-hole thermodynamics and evaporation, stability under perturbations, and dynamical observables such as quasinormal modes and grey-body factors. Optical phenomena, including black-hole shadows and lensing, as well as implications for gravitational-wave astronomy, have also been considered, linking the formal framework to potential observational signatures.

A common pattern emerging from different constructions is the appearance of universal features: the replacement of the central singularity by a regular core and at the same time having an identical asymptotic behavior as the cases in GR, the existence of a critical parameter below which no horizon forms, the tendency of the Hawking temperature to vanish near extremal configurations, leading to cold and long-lived remnants, and the shrink of the shadow. Such qualitative similarities, despite differences in the details of the scale-setting procedure or the truncations employed, reinforce the robustness of the Asymptotic Safety scenario in addressing fundamental issues such as singularity resolution and black-hole evaporation.

From a phenomenological perspective, AS-inspired black holes often display only percent-level deviations from their classical counterparts in the most accessible observables, such as the fundamental quasinormal mode or the shadow radius. Nonetheless, more pronounced differences can appear in regimes of strong gravity, particularly in the spectrum of higher overtones, near-extremal thermodynamics, or in the long-time evolution of Hawking evaporation. These are precisely the domains where future observational facilities, such as the next generation of gravitational-wave interferometers (Einstein Telescope, LISA) and increasingly precise imaging by the Event Horizon Telescope, could probe quantum-gravity imprints with sufficient sensitivity.

At the same time, significant challenges remain. A central open issue is the identification of the RG scale in curved spacetimes, which is not unique and may influence physical predictions, in fact, quantitative results depend on the chosen cutoff of the effective action, raising questions about the universality of the framework. Furthermore, the chosen truncation of the effective action can limit the type of solutions and their properties (see \cite{DelPorro:2025wts} for a recent exploration with Einstein-Weyl truncation).

In summary, the study of black holes in Asymptotic Safety has matured into a rich and active research direction that bridges quantum gravity, classical relativity, and astrophysical phenomenology. Continued progress in this area promises not only to refine our understanding of quantum-corrected black holes but also to open new avenues for testing the Asymptotic Safety scenario through current and future observational channels.


		\section*{Funding}
 This research received no external funding.
 
		\section*{Institutional Review Board Statement}
Not applicable.

		\section*{Informed Consent Statement}
 Not applicable.

		\section*{Data Availability Statement}
Not applicable.
 
		\section*{Acknowledgments}
The author acknowledges R. Konoplya and A. Bonanno for useful discussions. The author is also grateful for the hospitality of the Institute of Physics at Silesian University where part of the work was carried out.

		\section*{Conflicts of Interest}
The authors declare no conflict of interest.

\section*{Use of AI and AI-Assisted Technologies}

No AI tools were utilized for this paper.

	\small
	\bibliographystyle{scilight}
	
	

\end{document}